\documentclass[prl,twocolumn,nofootinbib,showpacs,superscriptaddress]{revtex4-1}

\usepackage{amsfonts}
\usepackage{amsmath}
\usepackage{amssymb}
\usepackage{bm}
\usepackage{dcolumn}
\usepackage{graphicx}
\usepackage{graphics}
\usepackage[latin1]{inputenc}
\usepackage{latexsym}
\usepackage{rotating}
\usepackage{xspace} 
\usepackage[usenames]{color}
\usepackage{mathrsfs}
\usepackage{subfigure}
\usepackage{multirow}

\usepackage{ulem}
\definecolor {darkgreen}{rgb}{0.2,0.7,0.2}

\newcommand\be{\begin{equation}}
\newcommand\ba{\begin{eqnarray}}
\newcommand\ee{\end{equation}}
\newcommand\ea{\end{eqnarray}}

\newcommand\bw{\begin{widetext}}
\newcommand\ew{\end{widetext}}

\newcommand{\nn}{\nonumber}

\begin{document}
\title{Nature Abhors a Circle}

\author{Nicholas Loutrel}
\affiliation{Department of Physics, Princeton University, Princeton, NJ 08544, USA.}

\author{Samuel Liebersbach}
\affiliation{eXtreme Gravity Institute, Department of Physics, Montana State University, Bozeman, MT 59717, USA.}

\author{Nicol\'as Yunes}
\affiliation{eXtreme Gravity Institute, Department of Physics, Montana State University, Bozeman, MT 59717, USA.}

\author{Neil Cornish}
\affiliation{eXtreme Gravity Institute, Department of Physics, Montana State University, Bozeman, MT 59717, USA.}

\date{\today}

\begin{abstract} 

The loss of orbital energy and angular momentum to gravitational waves produced in a binary inspiral forces the orbital eccentricity to adiabatically evolve and oscillate. 
For comparable-mass binaries, the osculating eccentricity is thought to decrease monotonically in the inspiral.
Contrary to this, we here show that, once the osculating eccentricity is small enough, radiation reaction forces it to grow secularly before the binary reaches the last stable orbit. 
We explore this behavior, its physical interpretation and consequences, and its potential impact on future gravitational wave observations.

\end{abstract}



\maketitle


\noindent \textit{Introduction.}~The recent detections of gravitational waves from compact binaries has provided invaluable information about the dynamical, strong field regime of gravity and the astrophysical processes that drive these systems to coalescence~\cite{GW150914, TheLIGOScientific:2017qsa, Yunes:2016jcc, Palmese:2017yhz}. While these observations have placed significant constraints on the merger rate of compact objects, their formation scenario remains unclear. One possibility is that these binaries formed from a co-evolving stellar binary, whereby two massive main sequence stars become either neutron stars or black holes through stellar evolution processes~\cite{Belczynski:2016obo}. By the time the gravitational waves emitted by these binaries enter the sensitivity band of ground-based detectors, their orbital eccentricity is expected to be very small. On the other hand, a non-negligible fraction of the systems may form in dense stellar environments, such as galactic nuclei and globular clusters~\cite{2009MNRAS.395.2127O, Antonini:2015zsa, Park:2017zgj,Samsing:2017xmd}. Dynamical friction forces the most compact objects to fall toward the gravitational center of these systems, where multi-body encounters can create binaries with a wide range of orbital eccentricities. Thus, extracting the orbital eccentricity from future gravitational wave observations may be a powerful tool to discriminate between formation channels~\cite{Nishizawa:2016jji}. 

In the \textit{post-Newtonian} (PN) approximation one solves the Einstein equations assuming small velocities/weak fields, with \emph{radiation reaction}, i.e. the back reaction of gravitational waves on the orbital dynamics of the binary that leads to a decaying orbit, typically included through an \emph{averaged balance law scheme}~\cite{Isaacson:1968gw}. The idea is that the averaged rate of change of the orbital binding energy and angular momentum must be balanced by the averaged rate at which gravitational waves carry energy and angular momentum away from the system. Since radiation reaction causes \textit{secular} changes in the orbital dynamics on timescales much longer than the orbital timescale, one then averages the gravitational wave fluxes over the orbital timescale~\cite{Isaacson:1968gw} before solving the balance law.

A more accurate picture of the inspiral and coalescence of binary systems can be obtained through the radiation-reaction force, i.e.~the force derived from the emission of gravitational waves that forces the orbit to decay. At leading PN order, the relative acceleration between two bodies is  $\vec{a} = \vec{f}_{\rm N} + \vec{f}_{\rm 2.5PN}$, where $\vec{f}_{N} = - (G M/r^{2}) \,\vec{n}$ is the Newtonian gravitational force with $M$ the total mass of the binary, and $(r, \vec{n})$ the radial separation between the two bodies and its associated unit vector. The second term in the relative acceleration is the leading PN order, radiation-reaction force, given explicitly in Eqs.~(12.221)-(12.222) of~\cite{PW}. Using the method of osculating orbits~\cite{PW, Pound:2010pj, Damour:2004bz}, this equation can be solved perturbatively by allowing the usual constants of the Kepler problem (such as the orbital energy and angular momentum) to evolve in time on a radiation-reaction timescale. The differential equations governing the evolution of the orbital element are given explicitly in Eqs.~(12.223)-(12.224) of~\cite{PW}. The two-body problem then reduces to simultaneously solving the relative acceleration equation and the evolution equations for the orbital elements. 

In each of these methods, the averaged balance law method and the osculating orbits method, the notion of eccentricity takes a slightly different meaning from the usual Keplerian sense. For conservative (closed) Keplerian orbits, the eccentricity provides a direct measure of the ellipticity of a closed orbit, and is thus referred to as the \textit{orbital eccentricity}. Once dissipation is taken into account, in this case due to radiation reaction, the orbits no longer close and the concept of orbital eccentricity becomes \emph{ill defined}. In the osculating method described above, dissipation is treated perturbatively, and the orbital eccentricity is promoted to a function of time. We refer to this eccentricity as the \textit{osculating eccentricity}, since it is not a fixed constant associated with closed orbits. Both of the methods discussed above to solve the radiation reaction problem evolve the osculating eccentricity.

\begin{figure}[ht]
\includegraphics[clip=true, width=\columnwidth]{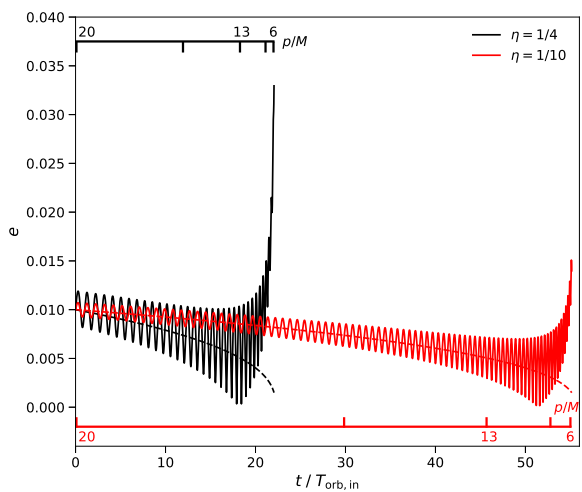}
\caption{\label{fig1} Temporal evolution of the osculating eccentricity relative to the initial orbital timescale $T_{\rm orb, in}$, obtained through the numerical evolution of the orbit averaged (dashed lines) and the osculating orbits (solid lines) equations. The scales display equal increments of the dimensionless semi-latus rectum $p/M$ for each system.}
\end{figure}

Although these two methods are distinct, they agree when one orbit averages the osculating orbit evolution, leading to the same secular changes to the orbit as what one finds with the averaged balance law method. The osculating orbit method, however, allows us to also study the effects of radiation reaction within an orbital timescale, which lead to \textit{oscillatory} modifications that vanish upon orbit-averaging. To illustrate this, Fig.~\ref{fig1} presents the temporal evolution of the osculating eccentricity calculated by numerically integrating the radiation-reaction equations given by the osculating orbits method (solid lines) and the orbit averaged approximation (dashed lines) for an equal-mass binary and a binary with mass ratio of $m_{2}/m_{1} \approx 0.127$. In all cases, we use the initial conditions $(p_{\rm in}, e_{\rm in}, \omega_{\rm in}, f_{\rm in}) = (20 G M/c^{2}, 10^{-2}, \pi, -\pi)$, where $p$ is the semi-latus rectum, $e$ is the osculating eccentricity, $\omega$ is the longitude of pericenter, and $f$ is the true anomaly, stopping the integrations when the system reaches the last stable orbit for a non-spinning test-particle $p = (2 GM/c^{2}) (3 + e)$. 

The evolution in the osculating method displays oscillatory behavior on the orbital timescale, while initially its secular change agrees with the orbit-averaged approximation. However, later in the evolution, roughly when the binary's semi-latus rectum is $p \approx 10$--$15 M$, corresponding to $p \approx 10^{3} \; {\rm{km}}$ for a binary with total mass $M = 60 M_{\odot}$, the osculating method produces a \emph{strong secular growth in the eccentricity}, which is opposite to what one obtains in the orbit-averaged approximation for comparable-mass binaries. This behavior seems counterintuitive for comparable-mass binaries, especially considering the wealth of literature on radiation reaction in the PN formalism, in which the eccentricity is always decreasing. However, it is important to remember that PN results are always computed within the orbit-averaged approximation. The radiation-reaction force is capturing effects beyond secular behavior that are not described in the orbit-averaged approximation.
\newline

\noindent \textit{Multiple Scale Analysis.}~To better understand this behavior, we consider a multiple scale analysis (MSA)~\cite{Bender, Mora:2003wt, Lincoln:1990ji, Pound:2007ti, Pound:2010pj} of the leading PN order, radiation-reaction equations, following~\cite{PW}. This analysis is valid provided $T_{\rm orb} \ll T_{RR}$, where $T_{RR} = | p / (dp/dt) |$ is the radiation reaction timescale and $T_{\rm orb}$ is the orbital timescale or simply the period of the orbit. Instead of using the variables $(p, e, \omega, t)$ with $f$ the dependent variable, we choose to work with $(p, A_{x}, A_{y}, t)$, where $(A_{x}, A_{y}) = (e \; {\rm cos}\omega, e \; {\rm sin}\omega)$ are the components of the Runge-Lenz vector, and with the orbital phase $\phi = f + \omega$ as the dependent variable. Working with these variables has the advantage of removing the $e^{-1}$ divergences in $(d\omega/df)$ and $(dt/df)$, as can be see in Eqs.~(12.223c) and~(12.224) of~\cite{PW}. With these variables, the osculating eccentricity can be easily reconstructed from $e = (A_{x}^{2} + A_{y}^{2})^{1/2}$.

Let us then define a few dimensionless parameters to simplify the evolution system. We let $\epsilon = (8 \eta/5) ({M}/{p^{\star}})^{5/2}$, $\textsf{p} = p/p^{\star}$ and $\textsf{t} = t/ ({p^{\star}}^{3}/ m)^{1/2}$~\cite{PW}, with $\eta = m_{1} m_{2}/M^{2}$ the symmetric mass ratio of the binary with component masses $m_{1}$ and $m_{2}$, and $p^{\star}$ 
a representative length scale of the system. 
%
%
%
%
Let us now carry out a multiple scale analysis by defining a ``fast'' variable $\phi$ and a ``slow'' variable $\tilde{\phi}$, i.e.~$\tilde{\phi} = \epsilon \; \phi$ with $\epsilon \ll 1$ the small parameter defined above, seeking solutions of the form $\mu^{a} = \mu_{0}^{a}(\tilde{\phi}, \phi) + \epsilon \, \mu_{1}^{a}(\tilde{\phi}, \phi) + {\cal{O}}(\epsilon^{2})$, $\textsf{t} = \epsilon^{-1} \textsf{t}_{-1}(\tilde{\phi}, \phi) + \textsf{t}_{0}(\tilde{\phi}, \phi) + {\cal{O}}(\epsilon)$,  where $\mu^{a} = (\textsf{p}, A_{x}, A_{y})$. At orders ${\cal{O}}(\epsilon^{0})$ and ${\cal{O}}(\epsilon)$, the equations become
\begin{equation}
\label{eq:msa-0}
\frac{\partial \mu_{0}^{a}}{\partial \phi} = 0\,,
\end{equation}
\begin{equation}
\label{eq:msa-1}
\frac{\partial \mu_{0}^{a}}{\partial \tilde{\phi}} + \frac{\partial \mu_{1}^{a}}{\partial \phi} = F^{a}(\phi, \mu_{0}^{b})\,,
\end{equation}
which we solve order by order by first splitting the solution into a secular and an oscillatory contribution $\mu_{\ell}^{a} = \mu_{\ell, {\rm sec}}^{a}(\tilde{\phi}) + \mu_{\ell, {\rm osc}}^{a}(\phi, \tilde{\phi})$. Equation~\eqref{eq:msa-0} then mandates that $\mu_{0}^{a} = \mu_{0,{\rm sec}}^{a}(\tilde{\phi})$, i.e.~the orbital elements are constants on conservative Keplerian ellipses. We then use that $\mu^{a}_{1}$ and $F^{a}$ are periodic in $\phi$ to  orbit average Eq.~\eqref{eq:msa-1}, integrating out any oscillatory effects, which then yields a differential equation for $\mu_{0, {\rm sec}}^{a}$ that we can solve. 

The procedure described above can be taken systematically to any order in $\epsilon$ by keeping higher-order terms in the expansion and using the lower-order in $\epsilon$ solutions. To leading order in $\epsilon$, the secular evolution of the orbital elements $\mu_{0, {\rm sec}}^{a}$ is identical to that found through the orbit-averaged form of the balance law; this can be combined to obtain the secular evolution of the osculating eccentricity, in agreement with Peters~\cite{Peters:1964zz}.
%
%
%
%
At next order in $\epsilon$, we can calculate $(A_{x,1}^{\rm sec}, A_{y,1}^{\rm sec})$ and $(A_{x,1}^{\rm osc}, A_{y,1}^{\rm osc})$; we do not provide the full expressions here as they are rather lengthy and unilluminating. 

We are here specifically interested in the evolution of the osculating eccentricity. Expanding its definition in terms of the norm of the Runge-Lenz vector, we find 
\begin{align}
\label{eq:e-msa}
e^{2} &= \left[ A_{x,0}(\tilde{\phi})^{2} + A_{y,0}(\tilde{\phi})^{2} \right] 
\nn \\
&+ 2 \epsilon \left[A_{x,0}(\tilde{\phi}) A_{x,1}(\phi, \tilde{\phi}) + A_{y,0}(\tilde{\phi}) A_{y,1}(\phi, \tilde{\phi})\right] 
\nn  \\
&+ \epsilon^{2} \left[A_{x,1}(\phi, \tilde{\phi})^{2} + A_{y,1}(\phi, \tilde{\phi})^{2} + 2 A_{x,0}(\tilde{\phi}) A_{x,2}(\phi,\tilde{\phi}) 
\right.
\nn \\
&\left.
+ 2 A_{y,0}(\tilde{\phi}) A_{y,2}(\phi,\tilde{\phi})\right]\,,
\end{align}
keeping terms up to ${\cal{O}}(\epsilon^{2})$. Several important features are present in this equation. First, notice that $A_{x,0} = {\cal{O}}(v^{0}/c^{0}) = A_{y,0}$, while $A_{x,1} = {\cal{O}}(v^{5}/c^{5}) = A_{y,1}$  and $A_{x,2} = {\cal{O}}(v^{10}/c^{10}) = A_{y,2}$. This is because each new order in $\epsilon$ is suppressed by the ratio of orbital timescale to the radiation-reaction timescale, which is of ${\cal{O}}(v^{5}/c^{5})$. Second, both $A_{x,0}$ and $A_{y,0}$ are linear in the eccentricity, while all higher-order terms are independent of the eccentricity. This is because to leading-order in $\epsilon$ the eccentricity $e = (A_{x,0}^{2} + A_{y,0}^{2})^{1/2}$. Third, although terms linear in $A_{x,1}$ and $A_{y,1}$ (in the second line of Eq.~\eqref{eq:e-msa}), or linear in $A_{x,2}$ and $A_{y,2}$ (in the third and fourth lines of Eq.~\eqref{eq:e-msa}) contain oscillatory contributions that average out on the orbital timescale, terms quadratic in $A_{x,1}$ and $A_{y,1}$ at ${\cal{O}}(\epsilon^{2})$ \emph{do not average out} and produce secular growth in the orbital eccentricity. Moreover, even the linear terms contain secular contributions $(A_{x,1}^{\rm sec}, A_{y,1}^{\rm sec})$ at ${\cal{O}}(\epsilon)$ that do not vanish and also contribute to the secular growth; however, these contributions are smaller than those coming from the third line in Eq.~\eqref{eq:e-msa}.  

We thus arrive at a possible physical and mathematical explanation for the secular growth in the eccentricity shown in Fig.~\ref{fig1}. The leading-order contribution to the orbital eccentricity, $(A_{x,0}^{2} +  A_{y,0}^{2})$, does indeed dominate for general initial eccentricities, leading to a monotonic decrease in time, as expected from the work of Peters~\cite{Peters:1964zz} and subsequent work at higher PN order~\cite{Arun:2009mc}. Eventually, however, this monotonic decrease forces the eccentricity to be small enough that the leading-order $(A_{x,0}^{2} +  A_{y,0}^{2})$ terms become smaller than terms higher-order in $\epsilon$, forcing the eccentricity to grow monotonically. This occurs when ${\cal{O}}(A_{i,0}^{2}) = {\cal{O}}(A_{i,0} A_{i,1}) = {\cal{O}}(A_{i,1}^{2})$ for any component of the Runge-Lenz vector, which translates to $e \sim v^{5}/c^{5}$ because $A_{i,0}^{2} \sim e^{2}$, $A_{i,0} A_{i,1} \sim e \; v^{5}/c^{5}$ and $A_{i,1}^{2} \sim v^{10}/c^{10}$. Indeed, we see in Fig.~\ref{fig1} that the secular growth starts when the osculating eccentricity has decayed to roughly $10^{-3}$, so inverting $e \sim v^{5}/c^{5}$ this would corresponds to a velocity of $v \sim e^{1/5} c \approx 0.25 c$, which corresponds to a semi-latus rectum of roughly $15 M$, matching the results shown in Fig.~\ref{fig1}. Therefore, \emph{when the osculating eccentricity becomes small enough, the radiation-reaction force generates a secular growth in the osculating eccentricity that indicates a break down of the orbit-averaged approximation.}
\newline

\begin{figure}[ht]
\includegraphics[clip=true, width=\columnwidth]{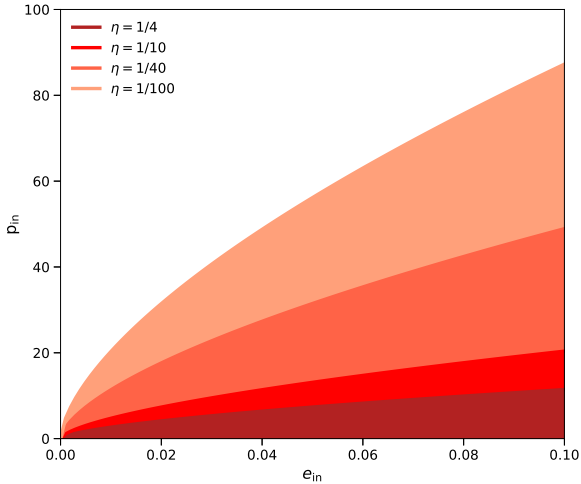}
\caption{\label{sep} Separatrix between secular decay and secular growth for binary systems of different symmetric mass ratio $\eta$ in the initial semi-latus rectum-initial eccentricity ($p_{\rm{in}}$--$e_{\rm in}$) plane. The shaded regions below each line corresponds to areas where secular growth does not occur.}
\end{figure}

\noindent \textit{Properties of Secular Growth.}~The secular growth of the osculating eccentricity is dependent on the mass ratio, as can be seen from Fig.~\ref{fig1}. A term of $N$th order in MSA scales with $\eta^{N}$, and since the growth enters at second order in MSA, it thus scales as $\eta^{2}$. In fact, in our analysis, one can make the mass ratio sufficiently small, such that the secular growth does not occur before the system reaches the last stable orbit. Let us then refine our approximation for the critical velocity at which the eccentricity of a binary switches from secular decay to secular growth, $e \sim v^{5}/c^{5}$, by including the mass-ratio dependence. Doing so, we find that the critical semi-latus rectum at which this occurs is 
\begin{equation}
p_{{\rm crit}} = \left(\frac{64 \; \eta}{ 5\; e_{\rm in}}\right)^{12/49} \left(\frac{p_{\rm{in}}}{M}\right)^{19/49} \left(1 - \frac{435}{2888} e_{\rm in}^{2}\right) M\,,
\end{equation}
where $p_{\rm{in}} = p(\tilde{\phi} = 0)$ is the initial semi-latus rectum. For the systems we consider in Fig.~\ref{fig1}, these correspond to $p_{{\rm crit}} = 13.12 M$ for the equal mass case and $p_{{\rm crit}} = 10.48 M$ for the $\eta = 1/10$ case, which agrees with Fig.~\ref{fig1}. If the critical separation is smaller than the separation at the last stable orbit, then the secular growth will not occur in the inspiral phase. Therefore, there is a \emph{separatrix} in the initial separation that divides the initial parameter space into regions where the growth will and will not occur, specifically
\begin{equation}
p_{\rm{in}, {\rm sep}} = 6^{49/19} \left(\frac{5 \; e_{\rm in}}{{64 \; \eta}}\right)^{12/19} \left(1 + \frac{21315}{54872} e_{\rm in}^2 \right) M\,.
\end{equation}
Figure~\ref{sep} shows this separatrix for different symmetric mass ratio, for a choice of $p^{\star} = M$ that corresponds to a choice of units. The shaded regions in this figure correspond to areas where secular growth does not occur. Observe that if the mass ratio is sufficiently small, or the initial eccentricity is sufficiently large, the growth does not occur before the last stable orbit.

The secular growth discovered here is \emph{not} an artifact of the PN approximation. The PN approximation is valid provided the orbital velocities are small, i.e.~$v/c = G M/(r c^{2}) \ll 1$, while the secular growth begins to occur roughly when $v/c \approx 1/4$--$1/3$. We have verified that the PN approximation is not the culprit of the secular growth by completing the equations of motion to 4.5PN order, specifically $\vec{f}_{\rm cons} = \vec{f}_{\rm 1PN} + \vec{f}_{\rm 2PN} + \vec{f}_{\rm 3PN} + \vec{f}_{\rm 4PN}^{\rm cons}$ for the conservative part of the force, and $\vec{f}_{\rm diss} = \vec{f}_{\rm 2.5PN} + \vec{f}_{\rm 3.5PN} + \vec{f}_{\rm 4PN}^{\rm tail} + \vec{f}_{\rm 4.5PN}$ for the radiation reaction force~\cite{Konigsdorffer:2003ue, Marchand:2017pir, Marchand:2016vox, Will:2016pgm}. The conservative PN corrections introduce large oscillations in the osculating eccentricity, which ultimately force the secular growth to occur earlier in the inspiral than in the leading PN order case.

The secular growth is also \emph{not} an artifact of the multiple scale expansion or of a gauge choice. Multiple scale analysis is valid when the ratio of the timescale is less than the value of the expansion parameter, i.e.~$T_{\rm orb}/T_{\rm RR} \ll \epsilon$. We have verified numerically that this inequality is satisfied in the entire domain for the systems in Fig.~\ref{fig1}, reaching its worst at the last stable orbit where $T_{\rm orb}/T_{\rm RR} \sim 0.2$ and $\epsilon = 0.4$ for equal-mass binaries and a choice of $p^{\star} = M$. A non-averaged multiple-scale analysis, as that of Eqs.~\eqref{eq:msa-0} and~\eqref{eq:msa-1}, is gauge dependent~\cite{MTW,PW}, but we have verified, both numerically and analytically, that there is not enough gauge freedom to remove the secular growth found here.
\newline

\noindent \textit{Comparison to Other Methods and Approximations.}~The secular growth of osculating eccentricity has been observed in two other scenarios. One of them is extreme mass ratio inspirals (EMRIs), where a small compact objects spirals into a supermassive black hole~\cite{Cutler:1994pb, Tanaka:1993pu, Apostolatos:1993nu}. EMRI studies are based on a finite expansion about small mass ratio, which allows for solutions to the Einstein equation that are valid to all PN orders. In this formalism, one can show, \emph{at linear order in the mass ratio}, that the orbital eccentricity grows as the small object approaches the separatrix between stable and plunging orbits. Our analysis cannot recover this result because we work to a finite PN order, which converges poorly in the extreme mass-ratio limit, and thus, is not valid at the separatrix. Our finite PN order analysis, however, does find a secular growth \emph{at second order in the mass ratio}, which should also be present in EMRI studies if one were to carry out calculations with the second-order self-force~\cite{Pound:2005fs, Pound:2007ti, Pound:2007th}.

The other scenario in which secular growth of the eccentricity has been previously observed is in comparable-mass inspirals within the PN formalism~\cite{Lincoln:1990ji}. Lincoln \& Will studied the inspiral of binary systems using the osculating method, but with the perturbing force $\vec{f}_{\rm 1PN} + \vec{f}_{\rm 2PN} + \vec{f}_{\rm 2.5PN}$. Using a two time-scale analysis, they computed the leading order secular (orbit averaged) effects at each PN order in the perturbing force, as well as the oscillatory corrections from $\vec{f}_{\rm 1PN}$. While they did see a growth in eccentricity, the latter also exhibited oscillations with significantly higher amplitude than what we find in Fig.~\ref{fig1}. Furthermore, Lincoln \& Will also found that the system enters a ``state of perpetual apastron," where the true anomaly becomes $f=\pi$ for the remainder of the evolution. As a result of this, zero Keplerian eccentricity no longer corresponded to a circular orbit in their analysis~\cite{Lincoln:1990ji, PW}, with the PN corrections to the eccentricity given in Eq.~(3.5) therein. 

The cause of the growth seen by Lincoln \& Will can be seen from Eq.~(3.14) in~\cite{Lincoln:1990ji}, which consists of two terms: a leading PN order term and a 1PN order correction. At large separation, the Keplerian eccentricity decreases monotonically, as controlled by the leading PN order term under orbit-averaged radiation-reaction. As the separation decreases throughout the inspiral, the 1PN correction grows until eventually this term becomes larger than the leading PN order one late in the inspiral, producing the growth seen by Lincoln \& Will. This implies that their results are not due to higher order effects in a MSA, but are rather due to an admixture of conservative and dissipative effects at various PN orders. Because of the role of conservative effects in their secular growth, the latter can actually be eliminated by a redefinition of the Keplerian eccentricity.

The effect we have found is similar to that seen by Lincoln \& Will (in that it is also a growth of an eccentricity parameter), but our analysis shows that this growth is due to the non-averaged effect of the dissipative radiation-reaction force, and not to uncontrolled post-Newtonian remainders or an issue of the parametrization. Indeed, as we explained before, we have verified that this growth remains even when one includes higher PN order terms in both the conservative and the dissipative sector. Moreover, our multiple scale analysis has shown analytically that the growth arises due to nonlinear effects in the dissipative radiation-reaction force. Our results, however, do confirm the interpretation of Lincoln \& Will that when the growth starts the system enters a state of perpetual periastron, indicating that the growth is a feature of the binary system transitioning from an inspiraling elliptical state into a quasi-circular one. However, our perturbing force is purely odd under time reversal, and thus, redefinitions of the conservative eccentricity, like the one described above, are not possible. The growth we describe here is thus a fundamental end state of the inspiral under the effect of radiation reaction.

\begin{figure}[ht]
\includegraphics[clip=true, width=\columnwidth]{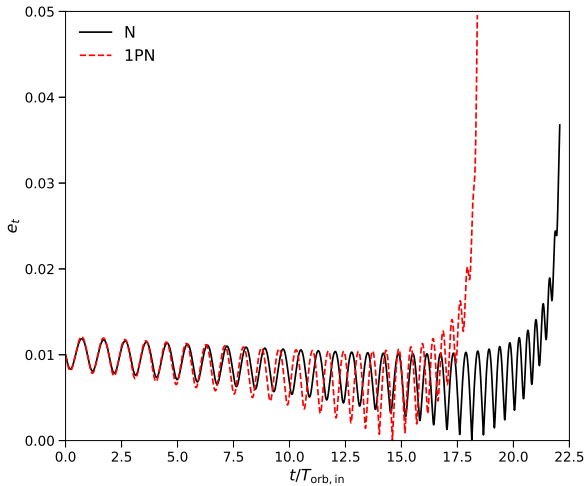}
\caption{\label{qk} Time evolution of the quasi-Keplerian~\cite{zbMATH03938612, zbMATH04001537} time eccentricity $e_{t}$ at relative Newtonian order (solid line) and 1PN order (dashed line) using the method of variation of constants.}
\end{figure}

In order to address the meaning of eccentricity at higher PN orders, we have also worked in the quasi-Keplerian (QK) formalism of Damour and Deruelle~\cite{zbMATH03938612, zbMATH04001537}, in which circular orbits do correspond to zero QK eccentricities. The QK formalism provides a PN accurate description of the conservative dynamics of the binary without relying on the osculating approximation. Dissipation due to the radiation reaction force has been calculated within the QK formalism using the method of variation of constants~\cite{Damour:2004bz, Konigsdorffer:2006zt}, which is a variation on the method of osculating orbits. As expected, when orbit-averaging radiation reaction and applying the averaged balance laws, the QK eccentricities do monotonically decrease as the binary inspirals~\cite{Arun:2009mc}, but what if one does not use orbit-averaging? We have numerically investigated whether the secular growth exists within the QK formalism to relative 1PN order, with the results for the time eccentricity $e_{t}$ as a function of time displayed in Fig~\ref{qk}. The Newtonian evolution (solid line) is the same as Fig.~\ref{fig1}, while the initial conditions for the 1PN evolution (dashed line) are chosen such that the mean motion and time eccentricity are the same as in Fig.~\ref{fig1}.

Although the binary coalesces faster, the eccentricity still grows at 1PN order, occurring at $\sim 81\%$ of the total evolution time ($t_{\rm growth} \approx 14.5 T_{\rm orb, in}$ while $t_{f} \approx 18 T_{\rm orb, in}$). This is not significantly different from the Newtonian case, where the growth occurs at $t_{\rm growth} \approx 18 T_{\rm orb, in}$, which is $\sim 82\%$ of $t_{\rm f} \approx 22 T_{\rm orb, in}$. This should not be unexpected; the growth is a result of second order terms in the MSA, which scale like 5PN corrections in Eq.~\eqref{eq:e-msa}. The 3.5PN radiation reaction effects on the growth will be suppressed by $(v/c)^{2}$ relative to these. However, the inclusion of 3.5PN radiation reaction effects does enhance the final eccentricity at the end of the growth: at 1PN order, $e_{t}(t = t_{\rm f}) \approx 0.05$, corresponding to a $\sim 43\%$ increase relative to the Newtonian case. All of this indicates that the growth is not a result of a particular choice for conservative orbital parameterization.

\begin{figure}[ht]
\includegraphics[clip=true, width=\columnwidth]{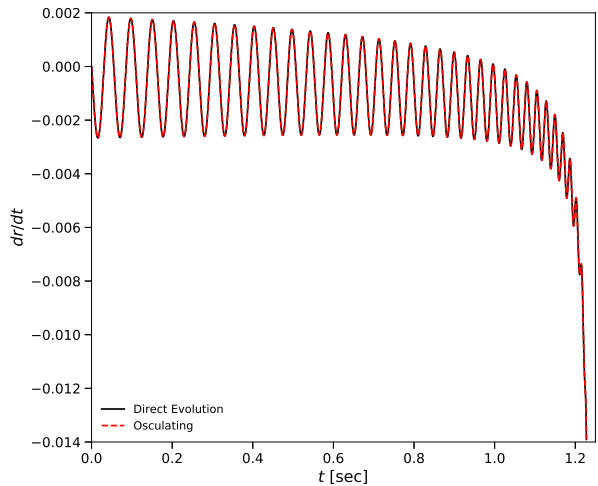}
\caption{\label{rdot} Radial velocity of a binary system as calculated through the direct evolution method (solid line) and the osculating approximation (dashed line).}
\end{figure}

The growth in the osculating eccentricity is not easy to extract from numerical relativity (NR) simulations. A different notion of ``eccentricity" that can be extracted from NR simulations is the \textit{coordinate eccentricity}, which is read off from the envelope of the radial velocity $\dot{r}$. To illustrate this, it is useful to consider an alternative method of solving the leading PN order radiation reaction problem. Rather than assume a Keplerian parametrization for the orbit and use the method of osculating orbits, one could simply evolve the relative coordinates of the binary directly using the equations of motion, a method we will refer to as \textit{direct evolution}. We plot $\dot{r}$ in Fig.~\ref{rdot} using this method for the same binary as Fig.~\ref{fig1}, and compare this to the reconstructed $\dot{r}$ in the osculating approximation, using the Keplerian expression $\dot{r} = -(M/p)^{1/2} (\vec{A} \cdot \vec{\lambda})$, where $\vec{\lambda} = (-{\rm sin}\phi, {\rm cos}\phi, 0)$ is the azimuthal unit vector of the orbit.
Notice first that the $\dot{r}$ in the osculating approximation agrees very well with the $\dot{r}$ in the direct evolution, with differences of the order of numerical error. This indicates that these two methods are equivalent descriptions of the binary. Second, the envelope of $\dot{r}$, and thus the coordinate eccentricity, decreases monotonically down to the last stable orbit, in contrast to the secular growth that is seen in the osculating eccentricity. 


How is it possible that two different measures of eccentricity, the osculating eccentricity and the coordinate eccentricity, give two different results? Eccentricity is not a gauge-invariant quantity, and it does not have a direct physical interpretation. The \textit{osculating} eccentricity is a time-dependent parameter in a specific orbital parametrization, which requires one be able to solve the conservative part of the two body problem analytically. On the other hand, the \textit{coordinate} eccentricity is a quantity that is reconstructed from the trajectories of the components of the binary, and thus, depends on the coordinate system one chooses to perform calculations therein. NR simulations do not have a well defined osculating approximation due to the non-linearity of the problem. Typically, statements of eccentricity in NR simulations are made through particular measures of coordinate eccentricity~\cite{Boyle:2007ft, Pfeiffer:2007yz, Tichy:2010qa}, although there has been some work on applying PN orbital parameterizations a posteriori to NR simulations~\cite{Buonanno:2010yk, Lewis:2016lgx}. The question posed above is, thus, difficult to answer in the context of NR simulations.

However, the measure of coordinate eccentricity can be applied to the PN two-body problem considered here. A rough sense of coordinate eccentricity can be obtained from the oscillations of the radial velocity (see Fig.~\ref{rdot}). We have here verified that the trajectories of the binary are consistent between the osculating approximation and the direct evolution, and the growth in the osculating eccentricity is equivalent to the decreasing coordinate eccentricity. Thus, there is no inconsistency between the secular growth seen in PN/EMRI calculations and the lack of growth in NR simulations. In fact, one could search for the secular growth in NR waveforms by performing a match calculation between them and PN waveforms in the orbit-averaged and the osculating approximations, provided numerical uncertainties are sufficiently under control. If the growth is in NR waveforms, then the match against osculating waveforms will be higher than the match against orbit-averaged waveforms. We explore comparisons between osculating and coordinate notions of eccentricity in more detail in~\cite{Loutrel:2018ydu}.
\newline

\begin{figure}[ht]
\includegraphics[clip=true, width=\columnwidth]{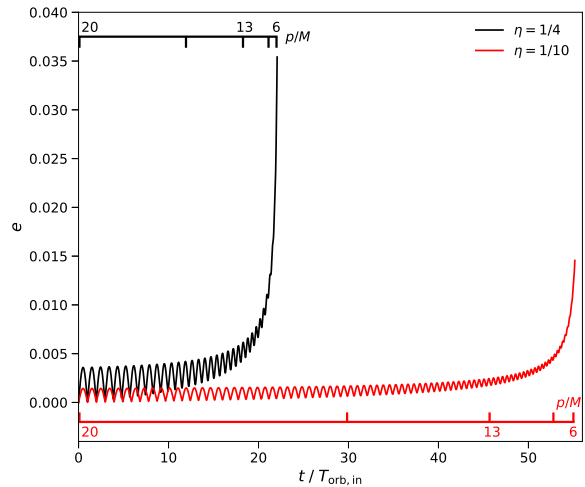}
\caption{\label{zero-ecc} Temporal evolution of the eccentricity for approximately initially circular binaries $(e_{\rm in} \ll 1)$, obtained through the numerical evolution of the the osculating orbits equations.}
\end{figure}

\noindent \textit{Discussion.}~Physically, the trajectories of the binary are becoming more and more quasi-circular as the binary inspirals, i.e. the oscillations in $\dot{r}(t)$ and $r(t)$ decrease, and the coordinate eccentricity decays, as the binary approaches the last stable orbit. However, this does not mean that the osculating eccentricity also decays; instead it tends to grow close to the last stable orbit. In fact, a system with small osculating eccentricity initially $e_{\rm in} \ll 1$ \textit{will not} remain at vanishingly small osculating eccentricity through the late inspiral, as can be seen in Fig.~\ref{zero-ecc}, which shows evolutions for the same systems as those presented in Fig.~\ref{fig1} but starting with small initial eccentricity. This secular growth is in stark contrast to what one would infer using the orbit-averaged approximation, where the orbit would continue to decay toward zero osculating eccentricity throughout the inspiral.
%


The secular growth of the osculating eccentricity does have an important observational implication for gravitational wave physics. The osculating eccentricity is actually the parameter that enters PN and EMRI waveform models, and thus, its growth enhances the harmonic content of the gravitational wave signal. Although a purely quasi-circular binary emits gravitational waves at twice its orbital frequency, a binary with non-vanishing osculating eccentricity emits gravitational waves at all harmonics of the orbital frequency.  The gravitational wave observations made by the advanced LIGO and Virgo detectors have not yet been sensitive enough to allow for a measurement of this harmonic content. But as these detectors are improved to achieve design sensitivity and third-generation detectors are built, a measurement of this harmonic content will become a reality. We have investigated the presence of eccentricity growth on GW observations from binary sources in~\cite{Loutrel:2018ydu}. We have found that the growth  is imprinted in the harmonic content of the waveform, and has a measurable effect on parameter estimation for sufficiently high signal-to-noise ratio sources. As ground-based detectors become more sensitive, and third generation detectors are built, the inclusion of this effect in waveform models will be important to limit biasing the recovered parameters of binary systems.

\acknowledgments We acknowledge support from the NSF CAREER grant PHY-1250636, NSF grant PHY-1607449, NASA grants NNX16AB98G and 80NSSC17M0041, the Simons Foundation, and the Canadian Institute For Advanced Research (CIFAR). We would like to thank Katerina Chatziioannou, Peter Diener, Davide Gerosa, Achamveedu Gopakumar, Scott Hughes, Sylvain Marsat, Amos Ori, Frans Pretorius, Leo Stein, Niels Warburton, Clifford Will, and Aaron Zimmerman for many useful comments and long discussions related to this work.

\bibliography{master}
\end{document}